\documentclass{elsart}

\usepackage{graphicx}
\usepackage{amssymb}
\begin{document}
\begin{frontmatter}
%

%
\title{Extraction of jet topology using three particle correlations}
%
%
\author{N. N. Ajitanand (for the PHENIX Collaboration) }
\address{Department of Chemistry, Stony Brook University, Stony Brook, NY 11794, USA}
\begin{abstract}
Recent theoretical studies  have indicated that the topological features 
of  away-side jet fragments can be significantly altered by  medium-induced 
modifications. The leading candidates resulting from such modifications are 
Mach Cones and deflected jets. We show that three particle correlations are
able to distinguish between these different modification scenarios. 
Initial results from an application of the method to Au+Au collisions at 
RHIC ($\sqrt{s_{NN}}=200$ GeV) are presented. 
\end{abstract}

\end{frontmatter}

\section{Introduction}
It is now generally accepted that in Au+Au  collisions at RHIC, a state of matter 
characterized as a strongly interacting, low viscosity fluid of quarks and 
gluons (termed sQGP) is created. In addition to the soft processes leading to  
to the formation of this medium, there are relatively rare hard parton-parton 
collisions.  The scattered partons may interact with the medium as they propagate 
through it, till they finally fragment into jet-like clusters. Since such interactions 
can modify jet fragmentation, jets provide a powerful probe of the medium, provided 
one can reliably extract the jet signal from the relatively large background which 
exists in RHIC collisions. Jet modification leading to shock wave induced conical 
flow (a ``sonic boom'') \cite{mach_cones} and a ``deflected-jet" induced by interactions 
between the propagating partons and the flowing medium \cite{Armesto_04} are leading 
candidates.

\section{Methodology and results: }
\subsection{Correlation Functions:} 
\begin{figure}[t]
\begin{center}
\includegraphics[width=1.\linewidth]{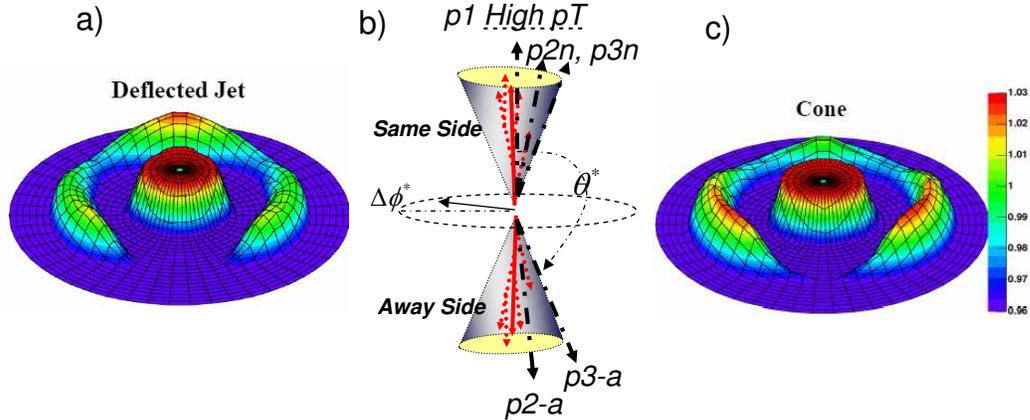}
\caption{(middle panel) Schematic illustration of the coordinate system used for 
the three-particle correlation analysis. The high-$p_T$ trigger particle serves as 
the near-side jet axis. $p2n$ and $p3n$ indicate associated particles on the near-side and 
$p2a$ and $p3a$ indicate associated particles on the away-side. (Left panel) Simulated three-particle 
correlation for a deflected away-side jet in the PHENIX acceptance. (Right panel) Same as left panel
but for a conical away-side jet.
}
\label{fig1}
\end{center}
\end{figure}

To study jet properties we use 
three-particle correlation functions built by combining a 
hadron in a specified high transverse momentum range ($2.5 < p_T < 4.0$ GeV/c) with two 
associated hadrons in an adjacent low transverse momentum range ($1.0 < p_T < 2.5$ GeV/c). 
The real triplets are obtained by combining all three particles from the same event. 
The mixed triplets are obtained by combining three particles from three different events. 
By transforming to the frame of reference in which the Z-axis is defined by the high $p_T$ 
particle momentum vector, a two dimensional correlation function is obtained 
in $\theta^*$, $\Delta\phi^*$ space where $\theta^*$ is the polar angle of one 
of the low $p_T$ particles and $\Delta\phi^*$ is the difference of azimuthal angles of 
the two associated particles (see Fig.~\ref{fig1}b). The three particle correlation functions 
can then be shown as polar plots in which $\theta^*$ is plotted along the radial axis 
and $\Delta\phi^*$ is plotted along the azimuthal axis (cf. Fig.\ref{fig1}b). 

\subsection{Simulation Results: } 
Simulated three-particle correlation surfaces are shown in 
Figs.~\ref{fig1}a and \ref{fig1}c  for two distinct away-side jet scenarios; 
(i) a ``deflected jet" in which the away-side jet axis is misaligned by $60^o$, and 
(ii) a ``Cherenkov or conical jet" in which the leading and away-side jet axes are aligned
but fragmentation is confined to a very thin hollow cone with a half angle of $60^o$. 
It is interesting to note here that these correlation surfaces reflect the removal of harmonic 
contributions but {\em do} include 2+1 three-particle combinatoric contributions. That is, 
2+1 correlations involving three particles, where two come from the jet/di-jet and the other 
from some other non-jet source. Nonetheless, both figures show rather clear distinguishing 
features for the two simulated scenarios considered.
\begin{figure}[t]
\includegraphics[width=1.\linewidth]{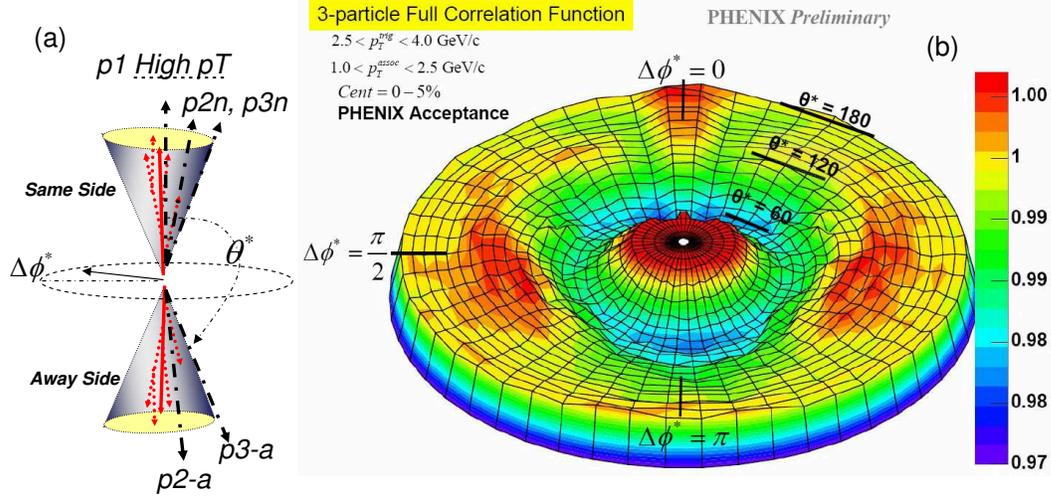}
\caption{(Color online)(a) Schematic illustration of coordinate system (same as Fig.~\ref{fig1}b).
(b) Full $\theta^*,\Delta\phi^*$ three particle correlation surface for charged hadrons 
detected in central (0-5\%) Au+Au collisions within the PHENIX acceptance. 
}
\label{3pc_cor_0-5}
\end{figure}

\subsection{Data Results: } 
The full three-particle correlation surface (no harmonic or 2+1 subtraction) 
obtained from data for the centrality selection 0-5\% in Au+Au collisions is 
shown in Fig. \ref{3pc_cor_0-5}b. A reminder of the coordinate frame used is given 
in Fig. \ref{3pc_cor_0-5}a. In this polar ($\theta^*,\Delta\phi^*$) representation, 
one can clearly see the near-side jet correlations which are expected at the center of the  
correlation surface ($\theta^*=0$). One can also see sizable correlations in $\Delta\phi^*$ for 
$\theta^* \sim 120^o$, albeit with acceptance losses especially in the region 
about ($\Delta\phi^* \sim 180^0$). Even without combinatoric background subtraction, it 
is apparent that the away-side jet shows a shape which is significantly different 
from that expected from a normal jet (ie. enhanced correlations for $\theta^* \sim 180^o$).
\begin{figure}[t]
\includegraphics[width=1.\linewidth]{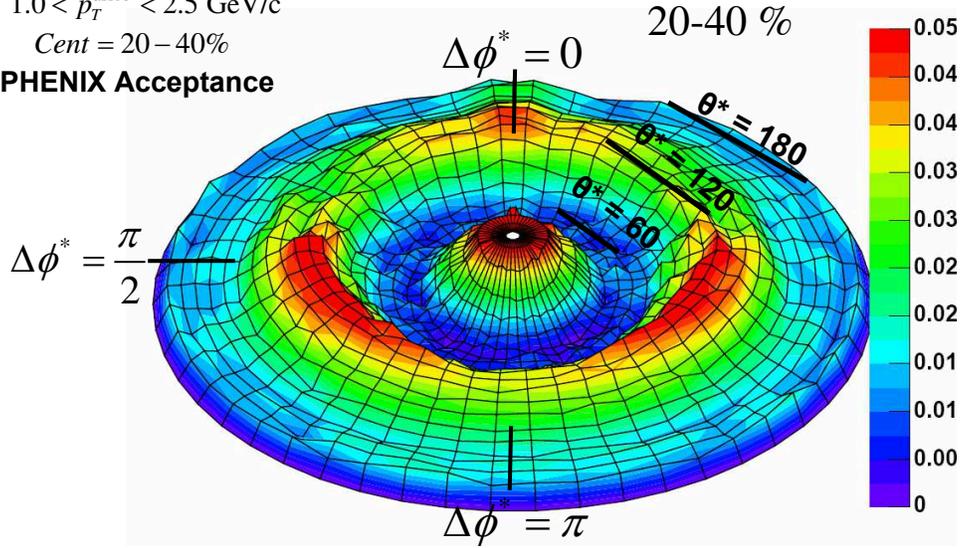}
\caption{(Color online) $v_2$ subtracted three particle correlation surface 
for charged hadrons detected in semi-central (20-40\%) Au+Au collisions 
within the PHENIX acceptance. 
}
\label{fig3}
\end{figure}

	Figure \ref{fig3} shows the resulting correlation surface 
(for centrality 20-40\%) after removal of the harmonic contributions 
via a two-dimensional equivalent of the ZYAM procedure outlined 
in Ref.~\cite{Ajitanand_05}.
In this case, the away-side jet correlations show a striking peak 
away from $180^o$ for most of $\Delta\phi^*$ albeit with inefficiencies 
associated with the PHENIX acceptance. Figs. \ref{3pc_cor_0-5}b and \ref{fig3} 
provide compelling evidence for strong modification of the away-side jet. 
	
	The correlations shown in Figs.~\ref{3pc_cor_0-5}a and \ref{fig3} are quite 
suggestive. However, a definitive and quantitative distinction of the mechanistic origin 
of the away-side jet modification requires the further step of removing the 
2+1 combinatoric contributions from Figs.~\ref{3pc_cor_0-5}a and \ref{fig3}. Given the potential 
utility of these correlation measurements as additional constraints for the viscosity 
and sound speed of hot QCD matter \cite{RLacey_sonic}, this step is being pursued 
vigorously. Suffice to say, systematic errors are currently being evaluated. 




\begin{thebibliography}{9}
%
%
\bibitem{mach_cones} H. St\"{o}cker, nucl-th/0406018; J. Casalderrey-Solana et al, hep-ph/0411315; 
B. Muller et al, Hep-ph/0503158; T. Renk et al, Hep-ph/0509036;
\bibitem{Armesto_04} Armesto,Salgado,Wiedemann hep-ph/0411341
%
%
\bibitem{Ajitanand_05} Ajitanand Alexander, Chung, Holzmann, Issah, Lacey, Shevel, Taranenko P. 
                  Danielewicz  Phys.Rev. C 72, 011902 (2005)
%
\bibitem{RLacey_sonic} Roy A. Lacey, nucl-ex/0608046; Roy A. Lacey, et al, nucl-ex/0609025
%
%
\end{thebibliography}
\end{document}